# Grain boundaries in minerals: atomic structure, phase transitions, and effect on strength of polycrystals

*Arslan B. Mazitov[1,2],[*], Artem R. Oganov[3]*

[1] *Moscow Institute of Physics and Technology, 9 Institutsky lane, Dolgoprudny 141700, Russia*
[2] *Dukhov Research Institute of Automatics (VNIIA), Moscow 127055, Russia*
[3] *Skolkovo Institute of Science and Technology, Skolkovo Innovation Center, 3 Nobel Street, Moscow 121205, Russia*

Corresponding Author

[*] Arslan B. Mazitov, e-mail: arslan.mazitov@phystech.edu

Grain boundaries (GBs) and interfaces in polycrystalline materials are significant research subjects in the field of materials science. Despite a more than 50-year history of their study, there are still many open questions. The main challenge in studying interfacial structures is the extreme complexity of their experimental and theoretical observation and description. The presence of phase-like states at grain boundaries called *complexions* requires even more effort in their study. Here, we demonstrate the effect of grain boundaries on the properties of polycrystalline minerals on the example of the Σ5(310)[001] grain boundary in periclase (MgO). Using the combination of extended evolutionary algorithm USPEX and modern machine-learning interatomic potentials, we explore the configuration space of the specified grain boundary and predict its possible phase-like states. In addition to the widely studied CSL-type structure, we found several stable GB complexions with various atomic densities at the boundary plane. Analysis of grain boundary excess volume of the structures revealed the successive stages of GB failure under the tensile applied in the normal direction of the boundary plane. Our results demonstrate that interfacial chemistry and structural diversity can be surprisingly rich even in seemingly simple and thoroughly investigated materials. The phenomena we observe here are not specific to MgO and should be general.

Grain boundaries and interfaces are known to strongly influence the mechanical and transport behavior of polycrystalline materials (1). Many efforts have been made in the last decades (2) to establish the relation between local interfacial structure and chemistry and various phenomena such as microstructure evolution (3–5), segregation (6), creep (7–10), fatigue (11–14), fracture (15, 16) and corrosion (17–19). According to the Gibbs phase rule, the intrinsic complexity of both homo- and heterophase interfaces requires at least five parameters to uniquely describe their misorientation and plane orientation (1). Thus, various approaches including Coincidence Site Lattice (CSL) and Displacement Shift Complete (DSC) lattice (1), dichromatic patterns and complexes (based on Shubnikov groups) (20–22), and Bollman's 0-lattice theory (23, 24) were introduced to provide this description. Recent experimental studies show that the chemical and structural diversity of grain boundaries can be even bigger. According to Hart (25), Dillon et al. (26, 27), and Cantwell et al. (28, 29), GB's can demonstrate phase-like behavior similarly to bulk crystals and undergo first-order transitions called *complexion* transitions at the same macroscopic crystallographic parameters. This makes the study of grain boundary structure and its relation to the properties of materials even more sophisticated.

Experimental observation of grain boundaries and complexion transitions on the atomic scale is extremely complicated since it requires a thorough sample growth and preparation followed by accurate *in situ* high-resolution microscopy analysis in ultrahigh vacuum (30, 31). On the other hand, theoretical simulations of GB's are relatively cheap since they are often carried out within classical molecular dynamics on empirical interatomic potentials, where CSL theory is used for creating the initial orientation of crystallites (32–37). However, this approach cannot provide a general description of all GB phases and complexion transitions for given GB crystallographic parameters, being limited only to manually prepared structural patterns. Furthermore, such calculations can suffer from inaccuracy of both CSL approximation and errors of empirical interatomic potentials. Despite the non-trivial nature of the problem, an effective method of grain boundary structure prediction based on evolutionary algorithm USPEX (38–40), was recently proposed and successfully applied to study of GB's phase behavior in elemental metals (41–44).

Here, we extend this method to grain boundaries in compounds, and boost its efficiency and reliability by using accurate machine learning interatomic potentials. USPEX was previously used to study bulk crystals at zero and finite temperatures (45), free surfaces (46), two-dimensional materials (47), epitaxial thin films (48), nanoclusters (49), and grain boundaries in single-component systems (41). The interface system in our method is represented as a combination of three parts: two bulk regions and an interfacial region located between them (Figure 1).

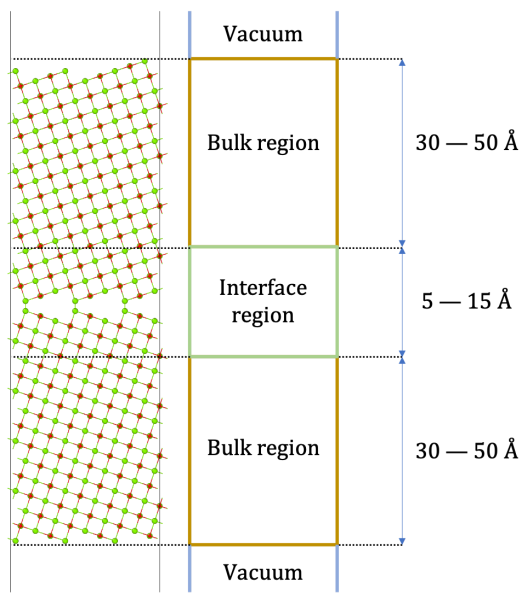

Figure 1. Interface model in the evolutionary algorithm. Structure optimization is performed in the interface region, while the bulk regions are frozen. Both bulk regions end up with free surfaces followed by vacuum layers that exclude the periodic replica interactions. Thickness of the interface region varies from 5 to 15 Å, while the bulk regions are usually 30 to 50 Å thick.

Our algorithm optimizes the structure of the interface keeping bulk regions unchanged. The first generation of individuals is produced with a topological random structure generator (50), while each subsequent one is created by variation operators using the best representatives of the previous generation as parents. There are four variation operators implemented in our method: heredity (1), softmutation (2), transmutation (3), and addition/deletion (4). The first one creates offspring from two randomly sliced parents by combining their fragments. The remaining operators manipulate the structure of a single parent by (1) shifting the atoms along with the softest vibrational modes (2), changing the chemical identity of some atoms (3), and adding (or deleting) the atoms in the system according to their coordination number (4). A certain percentage of random structures are included in each generation to ensure diversity of the population. After being created, all the structures are relaxed and ranked by fitness function based on their energy. This process is repeated iteratively until the set of the fittest individuals remains unchanged for a certain number of generations.

The evolutionary search typically requires hundreds or even thousands of structure relaxations, which should be performed with high accuracy to properly determine the ground state of the system.

In contrast to crystal structures, the atomic structures of computationally studied grain boundaries are usually larger by an order of magnitude, which makes their prediction with the usage of *ab initio* relaxation practically unfeasible. Previous studies of grain boundaries in simple metals (36, 41, 43, 44, 51) show that classical interatomic potentials in principle can be utilized for their simulation and structure prediction. However, the accuracy of these potentials is not always sufficient to properly determine the energy of the ground state, and the problem of low accuracy is particularly severe for grain boundaries. Moreover, the number of materials with available interatomic potentials is rather limited, while only a part of them is applicable for grain boundaries study. In this work, we replace classical potentials with a ML interatomic potential implemented in the MTP package (52), and utilize the two-stage relaxation scheme with two individually trained classes of potentials. The potentials of the first class were actively trained on relaxation trajectories of random crystal structures generated by USPEX. They can operate in a wide region of the phase space and perform the initial rough relaxation. The second class of potentials was trained on molecular dynamics trajectories of bulk supercells with the addition of point defects and random deformations of the unit cell. Since there were no structures with free surfaces and grain boundaries in such a training set, we extended it with GB structures randomly generated with USPEX and preliminarily relaxed with the potential of the first class. The resulting training set represent the relatively narrow region of the phase space, while the corresponding interatomic potentials turn out to be more accurate and can be used for a final structure relaxation. Data on energies, interatomic forces and stresses of the structures was obtained on density functional theory (DFT) level using VASP code (53, 54). Accuracy of the second-class interatomic potential in prediction of equilibrium lattice constant $a_0$ and elastic constants $C_{11}$, $C_{12}$, and $C_{44}$ is given in Table 1. For more information on active learning of ML interatomic potentials, the reader is referred to work (55).

|     | $a_0$, Å | $C_{11}$, GPa | $C_{12}$, GPa | $C_{44}$, GPa |
|-----|------|------|------|------|
| MTP | 4.252 | 285 | 101 | 154 |
| DFT | 4.249 | 272 | 90 | 143 |

Table 1. Comparison of equilibrium lattice constant $a_0$ and stiffness tensor components of MgO calculated with DFT and MTP.

We applied our method to study the atomic structure of Σ5(310)[001] grain boundaries in mineral periclase (MgO). This boundary is a symmetric tilt boundary resulting from the simultaneous rotation of two (310) surface slabs by an angle of 36.9° around the [001] axis. Our choice of this particular GB orientation was conditioned by a presence of a sufficient number of studies in the literature suitable for comparison of the results. Being probably the most studied metal-oxide material, MgO is commonly utilized as a model system in the analysis of more complex oxides with similar crystal structures. Grain boundaries are known to significantly affect various properties of polycrystalline metal oxides in a wide range of practically important applications, such as MOSFETs (56, 57), fuel cells (58, 59), gas sensors (60, 61), varistors (62), SQUIDs and high-Tc superconductors (63). According to a large number of experimental and theoretical studies (64–70), point defects in MgO tend to segregate at grain boundaries and diffuse along them, which makes the properties of polycrystalline MgO considerably depend on GB structure. The evolutionary search was carried out for up to 100 generations with 30 individuals in each generation. The initial population of 30 GB structures with 4-16 MgO units (up to 32 atoms) was produced by a topological random generator. The structures in subsequent generations were produced by heredity (40 %) and softmutation (20%), while the rest 30% was also produced randomly to diversify the population. The thickness of the grain boundary region was automatically adjusted according to the number of MgO units in the structure and the average atomic density of bulk MgO, while both bulk blocks were 50 Å thick. We also considered reconstructions of the grain boundary unit cell with a cell area up to 4 times larger than the original unit cell area. Relaxation and final energy calculation of the structures were performed in two stages with preliminary trained MTP interatomic potentials within the LAMMPS package (71). Each relaxation stage consisted of short finite temperature molecular dynamics run followed by conjugate gradient minimization, where only the atoms of the GB region were allowed to move. Finally, for each GB structure, we calculated the value of interface energy according to the formula

$$\gamma = \frac{1}{A}\left(E - 2E_s - N_{MgO}\mu_{MgO}\right)$$

where $E$ is the energy of whole atomic block with grain boundary, two bulk regions and two free (310) surfaces, $E_s$ is the (310) surface energy, $N_{MgO}$, $\mu_{MgO}$ are the number of MgO units in the structure and chemical potential of MgO, and $A$ is the area of the grain boundary.

Results of the evolutionary search are illustrated in Figure 2. We represented each structure as a point on the phase diagram in $(n, \gamma)$ space, where $n$ is the atomic density on the grain boundary plane. To derive $n$, we first calculate the number of atoms in the system $N$ and the number of atoms in one (310) plane of the MgO bulk region $N_{plane}^{bulk}$. Finally, we calculate $n$ as the ratio $\left(N \bmod N_{plane}^{bulk}\right)/N_{plane}^{bulk}$. Physical meaning of this quantity is the fraction of atoms from the ideal (310) plane located in the grain boundary plane. This approach, recently proposed by Zhu et al. (41), allows separating the structures with various atomic densities, considering them as grain boundary complexions. Our algorithm successfully found a widely known CSL-type structure (ID 587) with $n = 0$, which was extensively studied in previous works (33, 37, 66, 68) to be the most stable. Predicted value of its interface energy is $\gamma = 87 \; meV/Å^2$. Moreover, in addition to grain boundaries with standard atomic density ($n = 0$), various GB complexions with different atomic densities (ID 1570, ID 434, ID 252) were found. All these structures are essentially modifications of the ground state, which can be observed during grain boundary segregation and diffusion or complexion transitions. Interface energies of these structures $\gamma = 102 \; meV/Å^2$ (ID 1570), $\gamma = 105 \; meV/Å^2$ (ID 434), and $\gamma = 102 \; meV/Å^2$ (ID 252) are also fairly close to the ground state.

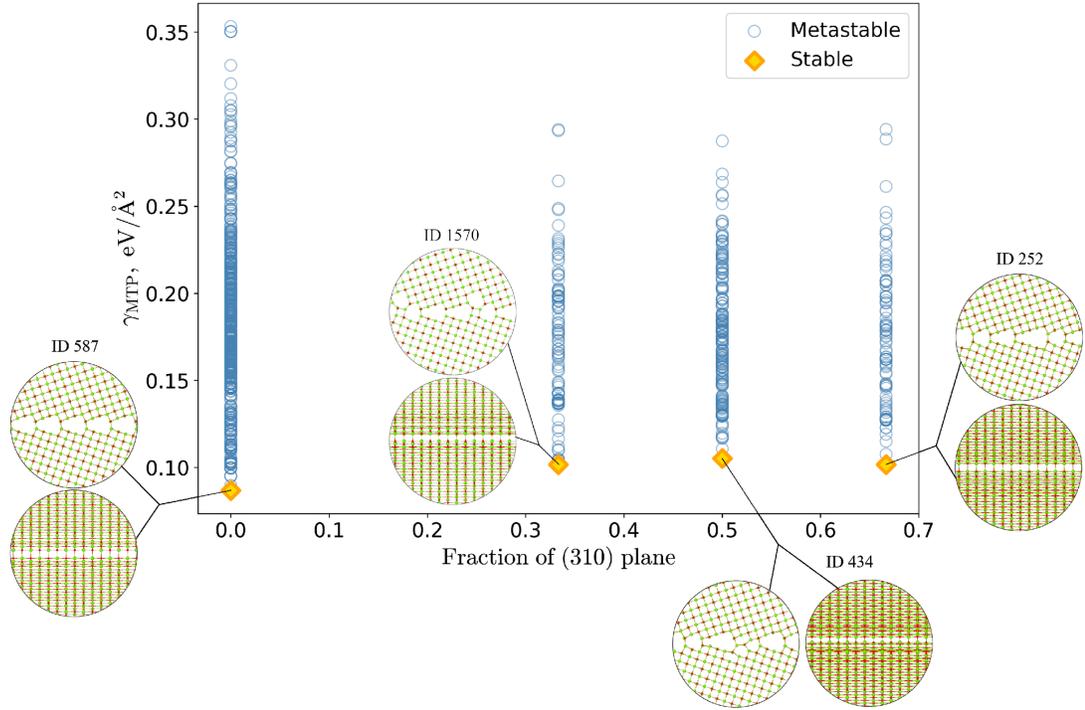

Figure 2. Phase diagram of the Σ5(310)[001] grain boundary structures in MgO. Metastable structures are represented with blue circles, stable ones - with orange diamonds. Front and side views of corresponding atomic configurations are given in insets, where magnesium atoms are marked with green spheres and oxygen atoms are marked with small red spheres. IDs of each presented structure are the IDs from evolutionary search. Ground state with $n = 0$ is a CSL-type structure (ID 587). Other newfound structures (ID 1570, ID 434, ID 252) are the modifications of the ground state with different atomic densities. They have relatively interface energies fairly close to the ground state, which may indicate possible complexion transitions inside the GB plane.

The accuracy of MTP interatomic potential in prediction of the interface energy was subsequently tested on a subset of the best GB structures from the evolutionary search. We selected 25 structures for each value of $n$ with various predicted $\gamma$ including both stable and unstable structures, and calculated their interface energies on DFT level with VASP. Results of the accuracy test are presented in Figure 3. MTP demonstrates an outstanding performance in prediction of the interface energy with a root-mean-squared error (RMSE) of 8.4 $meV/Å^2$ or 6.7 % with respect to mean value of $\gamma$ in a selected set of structures. Moreover, predicted interface energy of the ground state (ID 587) differ from DFT value only by 4.8 $meV/Å^2$, while the atomic configuration of the DFT ground state is almost indistinguishable from those predicted by MTP.

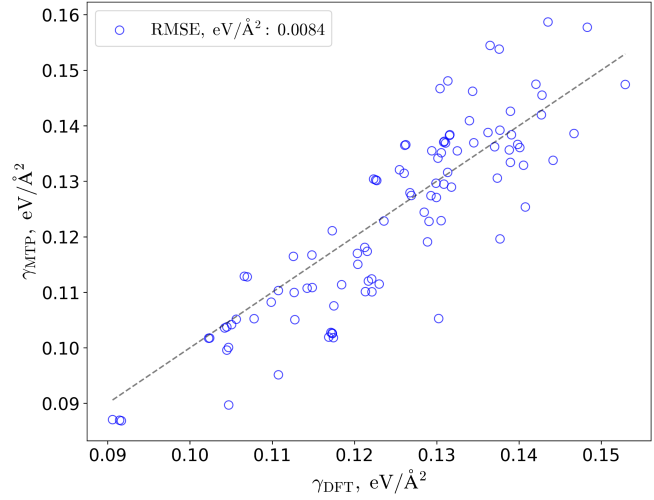

Figure 3. Comparison of the interface energy values predicted with MTP ($\gamma_{MTP}$) and calculated with VASP ($\gamma_{DFT}$). Each structure is represented as a blue circle, while a dashed gray line shows an ideal target prediction. The root-mean-squared error (RMSE) on MTP predictions is 8.4 $meV/Å^2$, or approximately 6.7 %.

Interface energies in Σ5(310)[001] GBs are rather ordinary for ionic crystals. In Table 2, we compare our results with typical values of grain boundary formation energies in different metals and minerals: copper, tungsten, diamond, tausonite, cubic zirconia and rutile. One can conclude that the formation energy depends more on the orientation of adjacent grains and the volume of CSL than on the nature of bonding in crystals. This result can be quite confusing, since there are no obvious reasons for grain boundaries in metal tungsten to be closer to covalently bonded diamond than to another metal like copper in terms of GB formation energy. The appearance of such surprising facts indicates that there are still many open questions in grain boundary physics and chemistry.

| Material | GB | $\gamma$, $meV/Å^2$ | Model | Ref. |
| --- | --- | --- | --- | --- |
| Copper (Cu) | Σ5(310)[001] | 56 | EAM | (35) |
| Tungsten (W) | Σ27(552)[11̄0] | 162 | DFT | (42) |
| Diamond (C) | Σ5(130)[011] | 172 | DFT | (72) |
| Tausonite ($SrTiO_3$) | Σ3(112)[1̄10] | 72 | DFT | (73, 74) |
| Cubic zirconia ($ZrO_2$) | Σ5(310)[001] | 45 | DFT | (75) |
| Rutile ($TiO_2$) | Σ13(221)[11̄0] | 47 | DFT | (76) |
| Periclase (MgO) | Σ5(310)[001] | 86 | MTP + DFT | This work |

Table 2. Comparison of the values of tilt grain boundary formation energies (γ) in various metals and minerals, calculated on density functional theory level (DFT), embedded atom model (EAM) and machine learning interatomic potential (MTP).

It is worth noting that all the GB structures found in the calculation have the considerable excess atomic volume, which is an important property of grain boundaries characterization. Large excess volume is usually considered proportional to the degree of segregation (77–79). In case of MgO, this tendency can be implicitly confirmed by experimental studies of grain boundary segregation (66, 80), where the cavities inside the GB plane act as the sinks for point defects. We represented the distribution of excess volume $\upsilon_{ex}$ in Figure 4. For each structure, we calculated the GB excess volume as $\upsilon_{ex} = 1/A(V - Nv_{MgO})$, where $V$ is the volume of arbitrary chosen region of the structure containing GB plane, $N$ is the number of atoms in this volume, and $v_{MgO}$ is the volume per atom in bulk MgO. Defined in this way, $\upsilon_{ex}$ represents the deviation of atomic volume in GB region from the bulk. Phase diagram constructed in ($\upsilon_{ex}$, γ) space allows to study the structural transitions resulting from application of stress $\sigma_{33}$ in the normal direction to the GB plane. Construction of the convex hull in such a diagram gives the information of stable phases at each value of $\sigma_{33}$, since it numerically equal to the negative slope of the convex hull sections: $\sigma_{33} = - \partial\gamma/\partial\upsilon_{ex}$. We note that since the GB structures were relaxed at zero pressure, all the deformations corresponding to applied stress are concentrated in GB plane and do not affect atoms in bulk regions.

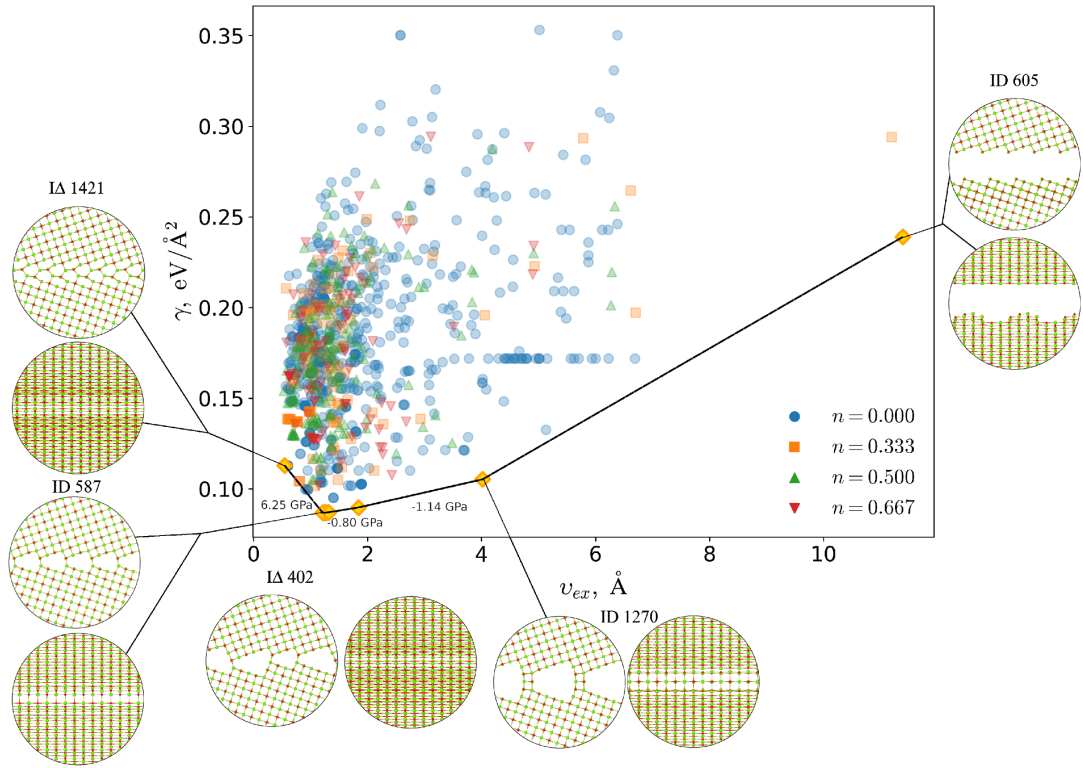

Figure 4. Phase diagram of the Σ5(310)[001] grain boundary structures in excess volume ($\upsilon_{ex}$) - interface energy (γ) space. Structures with different values of GB atomic density $n$ are represented with colored markers. Convex hull construction (black line) highlights the structures stabilized under stress applied in the normal direction to the GB plane (

$\sigma_{33}$), which is numerically equal to the negative slope of the convex hull sections. Corresponding values of $\sigma_{33}$ are given as label of convex hull sections in GPa units: positive sign for compression and negative sign for tension. Variation of the applied stress may lead to stabilization of dense ID 1421 structure on positive values of stress, or to gradual failure of GB structure (ID 605) with stabilization of intermediate configurations (ID 402, ID 1270) at $\sigma_{33} < 0$. Estimated value of the ultimate tensile strength is 1.14 GPa.

The values of $\sigma_{33} > 0$ represent the compression of the system and result in stabilization of dense ID 1421 structure, while the ground state (ID 587) is only stabilized at $\sigma_{33} \approx 0$. Change in the sign of $\sigma_{33}$ correspond to application of the tension along normal direction to the GB plane. It leads to gradual increase in excess volume of the structures (ID 402, ID 1270) and finally results in fracture of the grain boundary (ID 605). In other words, analysis of the stable GB phases in ($\upsilon_{ex}$, $\gamma$) space may reveal the fracture mechanics at the atomic scale and provide the thermodynamical description of related structural transitions. Particularly, one can estimate the value of tensile strength by calculating the slope of the convex hull section preceding the formation of a crack. The resulting value of Σ5(310)[001] grain boundary tensile strength is 1.14 GPa. This value is in a good correspondence with experimental observations of fracture strength in MgO bicrystals (81), which range from 0.05 to 0.3 GPa, depending on the temperature ($T = 300 - 1400$ K) and the average grain size ($D = 1 - 6$ mm). Possible deviations can be explained with a dramatic increase in fracture strength with a decrease in temperature and average grain size, and a specific character of particular GB orientation.

To sum up, here we demonstrate the application of our newly extended method to the prediction of the atomic structure of grain boundaries and interfaces. Utilizing the natural principles of Darwinian evolution, it is capable of automatically exploring grain boundary configurations space based only on knowledge of the structure, composition, and orientation of adjacent crystallites. The application of machine learning interatomic potentials for structure relaxations makes the study of interfacial structure possible for an arbitrary system of interest almost at *ab initio* level of accuracy. We tested our algorithm on the prediction of grain boundary complexions in MgO for Σ5(310)[001] orientation. To perform structure relaxation during the evolutionary search, two machine learning interatomic potentials were trained for two successive relaxation stages. The first one was trained on relaxation trajectories of randomly generated MgO crystal structures, and the second one - on molecular dynamics trajectories of MgO supercells with the addition of point defects and random deformations of the cell. Our results confirm that the well known CSL-type structure of Σ5(310)[001] GB is the ground state for this orientation, at least in zero-temperature case. In addition to the ground state with the same atomic density, several GB complexions with various atomic densities on the boundary plane were found. Moreover, the analysis of grain boundary excess volume revealed the mechanics of failure under applied stress on an atomic scale. The proposed methodology allows one to thoroughly investigate the configuration of arbitrary interfaces, which may significantly deepen our knowledge and understanding of this type of systems.

## Acknowledgments

A. B. M. thanks Russian Science Foundation (grant № 19-73-00237) for financial support in the development of the USPEX extension. A. R. O. thanks Russian Ministry of Science and Higher Education (grant № 2711.2020.2 to leading scientific schools) for financial support in study of phase-like transitions of grain boundaries under applied stress.

# References

1. A. P. Sutton, R. W. Balluffi, *Interfaces in Crystalline Materials* (Clarendon press, 1995).

2. J. H. Panchal, S. R. Kalidindi, D. L. McDowell, Key computational modeling issues in Integrated Computational Materials Engineering. *Comput. Des.* **45**, 4–25 (2013).

3. N. F. Mott, Slip at grain boundaries and grain growth in metals. *Proc. Phys. Soc.* **60**, 391–394 (1948).

4. P. A. Beck, P. R. Sperry, Strain induced grain boundary migration in high purity aluminum. *J. Appl. Phys.* **21**, 150–152 (1950).

5. J. E. Burke, D. Turnbull, Recrystallization and grain growth. *Prog. Met. Phys.* **3** (1952).

6. P. Lejček, M. Šob, V. Paidar, Interfacial segregation and grain boundary embrittlement: An overview and critical assessment of experimental data and calculated results. *Prog. Mater. Sci.* **87**, 83–139 (2017).

7. T. G. Langdon, Grain boundary sliding as a deformation mechanism during creep. *Philos. Mag.* **22**, 689–700 (1970).

8. R. Raj, M. F. Ashby, On grain boundary sliding and diffusional creep. *Metall. Trans.* **2**, 1113–1127 (1971).

9. T. Watanabe, Grain Boundary Sliding and Stress Concentration During Creep. *Metall. Trans. A, Phys. Metall. Mater. Sci.* **14 A**, 531–545 (1982).

10. Y. Chen, C. A. Schuh, Coble creep in heterogeneous materials: The role of grain boundary engineering. *Phys. Rev. B - Condens. Matter Mater. Phys.* **76**, 1–13 (2007).

11. K. Tanaka, Y. Akiniwa, Y. Nakai, R. P. Wei, Modelling of small fatigue crack growth interacting with grain boundary. *Eng. Fract. Mech.* **24**, 803–819 (1986).

12. M. D. Sangid, T. Ezaz, H. Sehitoglu, I. M. Robertson, Energy of slip transmission and nucleation at grain boundaries. *Acta Mater.* **59**, 283–296 (2011).

13. M. D. Sangid, The physics of fatigue crack initiation. *Int. J. Fatigue* **57**, 58–72 (2013).

14. W. D. Musinski, D. L. McDowell, Simulating the effect of grain boundaries on microstructurally small fatigue crack growth from a focused ion beam notch through a three-dimensional array of grains. *Acta Mater.* **112**, 20–39 (2016).

15. T. Watanabe, S. Tsurekawa, Control of brittleness and development of desirable mechanical properties in polycrystalline systems by grain boundary engineering. *Acta Mater.* **47**, 4171–4185 (1999).

16. T. Watanabe, S. Tsurekawa, Toughening of brittle materials by grain boundary engineering. *Mater. Sci. Eng. A* **387–389**, 447–455 (2004).

17. P. Lin, G. Palumbo, U. Erb, K. T. Aust, Influence of grain boundary character distribution on sensitization and intergranular corrosion of alloy 600. *Scr. Metall. Mater.* **33**, 1387–1392 (1995).

18. E. M. Lehockey, G. Palumbo, P. Lin, A. M. Brennenstuhl, On the relationship between grain boundary character distribution and intergranular corrosion. *Scr. Mater.* **36**, 1211–1218 (1997).

19. M. Shimada, H. Kokawa, Z. J. Wang, Y. S. Sato, I. Karibe, Optimization of grain boundary character distribution for intergranular corrosion resistant 304 stainless steel by twin-induced grain boundary engineering. *Acta Mater.* **50**, 2331–2341 (2002).

20. R. C. Pond, W. Bollmann, Symmetry and Interfacial Structure of Bicrystals. *Philos. Trans. R. Soc. London. Ser. A, Math. Phys. Sci.* **292**, 449–472 (1979).

21. R. C. Pond, D. S. Vlachavas, Bicrystallography. *Proc. R. Soc. London, Ser. A Math. Phys. Sci.* **386**, 95–143 (1983).

22. A. V. Shubnikov, V. A. Koptsik, *Symmetry in science and nature* (1972).

23. W. Bollmann, On the geometry of grain and phase boundaries. *Philos. Mag.* **16**, 383–399 (1967).

24. D. A. Smith, R. C. Pond, Bollmann's 0-Iattice theory; a geometrical approach to interface structure. *Int. Met. Rev.* **21**, 61–74 (1976).

25. E. W. Hart, H. Hu (ed.),. 155–170 (1972).

26. S. J. Dillon, M. P. Harmer, Multiple grain boundary transitions in ceramics: A case study of alumina. *Acta Mater.* **55**, 5247–5254 (2007).

27. S. J. Dillon, M. Tang, W. C. Carter, M. P. Harmer, Complexion: A new concept for kinetic engineering in materials science. *Acta Mater.* **55**, 6208–6218 (2007).

28. P. R. Cantwell, *et al.*, Grain boundary complexions. *Acta Mater.* **62**, 1–48 (2014).

29. P. R. Cantwell, S. Ma, S. A. Bojarski, G. S. Rohrer, M. P. Harmer, Expanding time–temperature-transformation (TTT) diagrams to interfaces: A new approach for grain boundary engineering. *Acta Mater.* (2016) https:/doi.org/10.1016/j.actamat.2016.01.010.

30. K. C. Chen, W. W. Wu, C. N. Liao, L. J. Chen, K. N. Tu, Observation of atomic diffusion at twin-modified grain boundaries in copper. *Science (80-. ).* **321**, 1066–1069 (2008).

31. P. R. Cantwell, *et al.*, Grain Boundary Complexion Transitions. *Annu. Rev. Mater. Res.* **50**, 465–492 (2020).

32. A. Béré, J. Chen, P. Ruterana, G. Nouet, A. Serra, A step associated with the Σ = 19 (2530) tilt boundary in GaN. *J. Phys. Condens. Matter* **14**, 12703–12708 (2002).

33. B. P. Uberuaga, X. M. Bai, Defects in rutile and anatase polymorphs of TiO2: Kinetics and thermodynamics near grain boundaries. *J. Phys. Condens. Matter* **23** (2011).

34. B. P. Uberuaga, X. M. Bai, P. P. Dholabhai, N. Moore, D. M. Duffy, Point defect-grain boundary interactions in MgO: An atomistic study. *J. Phys. Condens. Matter* **25** (2013).

35. T. Frolov, D. L. Olmsted, M. Asta, Y. Mishin, Structural phase transformations in metallic grain boundaries. *Nat. Commun.* **4**, 1897–1899 (2013).

36. T. Frolov, M. Asta, Y. Mishin, Segregation-induced phase transformations in grain

boundaries. *Phys. Rev. B - Condens. Matter Mater. Phys.* **92**, 1–5 (2015).

37. S. Fujii, T. Yokoi, M. Yoshiya, Atomistic mechanisms of thermal transport across symmetric tilt grain boundaries in MgO. *Acta Mater.* **171**, 154–162 (2019).

38. A. R. Oganov, C. W. Glass, Crystal structure prediction using ab initio evolutionary techniques: Principles and applications. *J. Chem. Phys.* **124**, 244704 (2006).

39. A. R. Oganov, A. O. Lyakhov, M. Valle, How evolutionary crystal structure prediction works-and why. *Acc. Chem. Res.* **44**, 227–237 (2011).

40. A. O. Lyakhov, A. R. Oganov, H. T. Stokes, Q. Zhu, New developments in evolutionary structure prediction algorithm USPEX. *Comp. Phys. Comm.* **184**, 1172–1182 (2013).

41. Q. Zhu, A. Samanta, B. Li, R. E. Rudd, T. Frolov, Predicting phase behavior of grain boundaries with evolutionary search and machine learning. *Nat. Commun.* **9**, 1–9 (2018).

42. T. Frolov, *et al.*, Grain boundary phases in bcc metals. *Nanoscale* **10**, 8253–8268 (2018).

43. T. Frolov, Q. Zhu, T. Oppelstrup, J. Marian, R. E. Rudd, Structures and transitions in bcc tungsten grain boundaries and their role in the absorption of point defects. *Acta Mater.* **159**, 123–134 (2018).

44. T. Meiners, T. Frolov, R. E. Rudd, G. Dehm, C. H. Liebscher, Observations of grain-boundary phase transformations in an elemental metal. *Nature* **579**, 375–378 (2020).

45. I. A. Kruglov, A. V. Yanilkin, Y. Propad, A. R. Oganov, Crystal structure prediction at finite temperatures (2021) (March 3, 2021).

46. Q. Zhu, L. Li, A. R. Oganov, P. B. Allen, Evolutionary method for predicting surface reconstructions with variable stoichiometry. *Phys. Rev. B* **87**, 195317 (2013).

47. X.-F. Zhou, *et al.*, Two-dimensional magnetic boron. *Phys. Rev. B* **93**, 85406 (2016).

48. A. B. Mazitov, A. R. Oganov, Evolutionary algorithm for prediction of the atomic structure of two-dimensional materials on substrates (2021).

49. S. V Lepeshkin, V. S. Baturin, Y. A. Uspenskii, A. R. Oganov, Method for simultaneous prediction of atomic structure and stability of nanoclusters in a wide area of compositions. *J. Phys. Chem. Lett.* **10**, 102–106 (2018).

50. P. V Bushlanov, V. A. Blatov, A. R. Oganov, Topology-based crystal structure generator. *Comput. Phys. Commun.* **236**, 1–7 (2019).

51. T. Frolov, M. Asta, Y. Mishin, Phase transformations at interfaces: Observations from atomistic modeling. *Curr. Opin. Solid State Mater. Sci.* (2016) https:/doi.org/10.1016/j.cossms.2016.05.003.

52. A. V Shapeev, Moment tensor potentials: A class of systematically improvable interatomic potentials. **14**, 1153–1173 (2016).

53. G. Kresse, J. Furthmuller, Efficient Iterative Schemes for Ab Initio Total-Energy Calculations Using a Plane-Wave Basis Set. *Phys. Rev. B.* **54**, 11169 (1996).

54. G. Kresse, D. Joubert, From Ultrasoft Pseudopotentials to the Projector Augmented-Wave

Method. *Phys. Rev. B.* **59**, 1758 (1999).

55. E. V. Podryabinkin, A. V. Shapeev, Active learning of linearly parametrized interatomic potentials. *Comput. Mater. Sci.* **140**, 171–180 (2017).

56. K. Kukli, *et al.*, Properties of hafnium oxide films grown by atomic layer deposition from hafnium tetraiodide and oxygen. *J. Appl. Phys.* **92**, 5698–5703 (2002).

57. V. Yanev, *et al.*, Tunneling atomic-force microscopy as a highly sensitive mapping tool for the characterization of film morphology in thin high-k dielectrics. *Appl. Phys. Lett.* **92** (2008).

58. J. Maier, Point-defect thermodynamics and size effects. *Solid State Ionics* **131**, 13–22 (2000).

59. T. Suzuki, I. Kosacki, H. U. Anderson, P. Colomban, Electrical Conductivity and Lattice Defects in Nanocrystalline Cerium Oxide Thin Films. *J. Am. Ceram. Soc.* **84**, 2007–2014 (2001).

60. I. Kosacki, C. M. Rouleau, P. F. Becher, J. Bentley, D. H. Lowndes, Nanoscale effects on the ionic conductivity in highly textured YSZ thin films. *Solid State Ionics* **176**, 1319–1326 (2005).

61. A. Dey, Semiconductor metal oxide gas sensors: A review. *Mater. Sci. Eng. B Solid-State Mater. Adv. Technol.* **229**, 206–217 (2018).

62. D. R. Clarke, Varistor Ceramics. *October* **502**, 485–502 (1999).

63. H. Hilgenkamp, J. Mannhart, Grain boundaries in high-Tc superconductors. *Rev. Mod. Phys.* **74**, 485–549 (2002).

64. D. M. Duffy, Grain boundaries in ionic crystals. *J. Phys. C Solid State Phys.* **19**, 4393–4412 (1986).

65. Y. Yan, *et al.*, Impurity-induced structural transformation of a mgo grain boundary. *Phys. Rev. Lett.* **81**, 3675–3678 (1998).

66. Y. Yan, M. F. Chisholm, G. Duscher, S. J. Pennycook, Atomic structure of a Ca-doped [001] tilt grain boundary in MgO. *J. Electron Microsc. (Tokyo).* **47**, 115–120 (1998).

67. S. C. Parker, D. J. Harris, Computer simulation of general grain boundaries in rocksalt oxides. *Phys. Rev. B - Condens. Matter Mater. Phys.* **60**, 2740–2746 (1999).

68. D. J. Harris, G. W. Watson, S. C. Parker, Atomistic simulation studies on the effect of pressure on diffusion at the MgO 410/[001] tilt grain boundary. *Phys. Rev. B - Condens. Matter Mater. Phys.* **64**, 134101 (2001).

69. J. H. Harding, Short-circuit diffusion in ceramics. *Interface Sci.* **11**, 81–90 (2003).

70. P. Wynblatt, G. S. Rohrer, F. Papillon, Grain boundary segregation in oxide ceramics. *J. Eur. Ceram. Soc.* **23**, 2841–2848 (2003).

71. S. Plimpton, Fast parallel algorithms for short-range molecular dynamics. *J. Comput. Phys.* **117**, 1–19 (1995).

72. M. Kohyama, Computational studies of grain boundaries in covalent materials. *Model. Simul. Mater. Sci. Eng.* **10**, R31 (2002).

73. A. L. S. Chua, N. A. Benedek, L. Chen, M. W. Finnis, A. P. Sutton, A genetic algorithm for predicting the structures of interfaces in multicomponent systems. *Nat. Mater.* **9**, 418–422 (2010).

74. S. Von Alfthan, *et al.*, The structure of grain boundaries in strontium titanate: Theory, simulation, and electron microscopy. *Annu. Rev. Mater. Res.* **40**, 557–599 (2010).

75. Z. Mao, S. B. Sinnott, E. C. Dickey, Ab initio calculations of pristine and doped zirconia Σ5 (310)/[001] tilt grain boundaries. *J. Am. Ceram. Soc.* **85**, 1594–1600 (2002).

76. G. Schusteritsch, *et al.*, Anataselike Grain Boundary Structure in Rutile Titanium Dioxide. *Nano Lett.* **21**, 2745–2751 (2021).

77. H. B. Aaron, G. F. Bolling, Free volume as a criterion for grain boundary models. *Surf. Sci.* **31**, 27–49 (1972).

78. T. Frolov, Y. Mishin, Thermodynamics of coherent interfaces under mechanical stresses. I. Theory. *Phys. Rev. B - Condens. Matter Mater. Phys.* **85**, 224106 (2012).

79. T. Frolov, Y. Mishin, Thermodynamics of coherent interfaces under mechanical stresses. II. Application to atomistic simulation of grain boundaries. *Phys. Rev. B - Condens. Matter Mater. Phys.* **85**, 224107 (2012).

80. Z. Wang, *et al.*, Atom-resolved imaging of ordered defect superstructures at individual grain boundaries. *Nature* **479**, 380–383 (2011).

81. R. C. Ku, T. L. Johnston, Fracture strength of MgO bicrystals. *Philos. Mag.* **9**, 231–247 (1964).